\documentclass[graybox]{svmult}

\usepackage{mathptmx}       
\usepackage{helvet}         
\usepackage{courier}        
\usepackage{type1cm}        
\usepackage{makeidx}         
\usepackage{graphicx}        
\usepackage{multicol}        
\usepackage[bottom]{footmisc}

\def\kmsec{km/s}
\def\lwt{$\langle t_\star \rangle_{\rm L}$}
\def\squareforqed{\hbox{\rlap{$\sqcap$}$\sqcup$}}                                                  
\def\sq{\ifmmode\squareforqed\else{\unskip\nobreak\hfil                                            
\penalty50\hskip1em\null\nobreak\hfil\squareforqed                                                 
\parfillskip=0pt\finalhyphendemerits=0\endgraf}\fi}  
\def\arcsec{\hbox{$^{\prime\prime}$}}  
\def\la{\mathrel{\mathchoice {\vcenter{\offinterlineskip\halign{\hfil                              
$\displaystyle##$\hfil\cr<\cr\sim\cr}}}                                                            
{\vcenter{\offinterlineskip\halign{\hfil$\textstyle##$\hfil\cr                                     
<\cr\sim\cr}}}                                                                                     
{\vcenter{\offinterlineskip\halign{\hfil$\scriptstyle##$\hfil\cr                                   
<\cr\sim\cr}}}                                                                                     
{\vcenter{\offinterlineskip\halign{\hfil$\scriptscriptstyle##$\hfil\cr                             
<\cr\sim\cr}}}}}                                                                                   
                         
\newcommand{\sbb}{mag/$\sq\arcsec$}

\makeindex             

\begin{document}

\title*{Pairwise dwarf galaxy formation and galaxy downsizing: some clues from extremely
  metal-poor Blue Compact Dwarf galaxies}
\titlerunning{Pairwise dwarf galaxy formation and galaxy downsizing}
\author{Polychronis Papaderos}
\institute{Centro de Astrof\'{\i}sica da Universidade do Porto
\email{papaderos@astro.up.pt}}
\maketitle

\vspace*{-2.5cm}
\abstract{Some of the extremely metal-poor Blue Compact Dwarf galaxies
(XBCDs) in the nearby universe form galaxy pairs with remarkably similar 
properties. This fact points to an intriguing degree of synchronicity 
in the formation history of these binary dwarf galaxies and raises 
the question as to whether some of them form and co-evolve pairwise  
(or in loose galaxy groups), experiencing recurrent mild interactions and minor
tidally induced star formation episodes throughout their evolution.
We argue that this hypothesis offers a promising conceptual framework 
for the exploration of the retarded previous evolution and 
recent dominant formation phase of XBCDs.}
\abstract*{Some of the extremely metal-poor Blue Compact Dwarf galaxies
(XBCDs) in the nearby universe form galaxy pairs with remarkably similar 
properties. This fact points to an intriguing degree of synchronicity 
in the formation history of these binary dwarf galaxies and raises 
the question as to whether some of them form and co-evolve pairwise  
(or in loose galaxy groups), experiencing recurrent mild interactions and minor
tidally induced star formation episodes throughout their evolution.
We argue that this hypothesis offers a promising conceptual framework 
for the exploration of the retarded previous evolution and 
recent dominant formation phase of XBCDs.}

\vspace*{-2mm}
\section{Introduction}
\label{sec:1}
Various lines of evidence indicate that some among the extremely metal-poor 
(12+log(O/H)$\leq$7.6) Blue Compact Dwarf (BCD) galaxies (hereafter XBCDs) 
in the nearby universe are cosmologically young objects, experiencing their 
dominant formation phase since $\sim$1--3 Gyr and containing a small, 
if any, mass fraction of ancient stars (see e.g. \cite{P08} for a discussion).
The origin of the recent rapid assembly of the stellar component of XBCDs is
generally ascribed to strong interactions or merging between gas-rich dwarf 
galaxies (DGs) having had undergone little previous star formation (SF) and 
chemical enrichment (see e.g. \cite{Pustilnik2008-IAU}).

However, the hypothesis of a chance encounter between two unrelated DGs does
not appear to be universally compelling.
This is because several XBCDs form galaxy pairs with a small linear (2\dots60 kpc)
and velocity separation (1\dots6 $\times$ their intrinsic velocity dispersion
of $\sigma\approx10$ \kmsec) and, most importantly, they are strikingly
similar with respect to their photometric structure, luminosity-weighted stellar age \lwt\ and
gas-phase metallicity.
One such example is the prototypical XBCD I\ Zw\ 18 and its C component. 
The former, after subtraction of its extended nebular halo, appears as almost 
a mirror image of the latter, with both galaxies being remarkably similar with
respect to their \lwt\ ($\simeq 10^8$ yr), isophotal radius (0.4--0.6 kpc at
26 \sbb) and exponential scale length $\alpha$ (100--130 pc) of their unresolved
\emph{stellar} host \cite{Papaderos2002-IZw18}. 
Another example is offered by the young XBCD candidates SBS\ 0335-052 E\&W, 
differing by only $\approx$0.2 dex in their gas-phase metallicity
(12+log(O/H)=7.1\dots7.3, \cite{Izotov2005-SBS0335W,Papaderos2006-SBS0335})
and by $\la$20\% in their $\alpha$ \cite{Papaderos1998-SBS0335}. 

Such properties point to an intriguing degree of synchronicity in the evolutionary
history of nearly equal-mass XBCDs, a fact that is difficult to reconcile with
the traditional hypothesis of a chance encounter between two unrelated DGs. 
They instead raise the question as to whether some XBCDs form and co-evolve as galaxy
pairs (or out of groups of low-mass baryonic entities) that possibly reside
within a common dark matter (DM) halo, and being exposed to perpetual mild 
mutual interactions throughout their evolution.
If so, a question that arises is whether this co-evolution process can,
under certain conditions, initially slow down the buildup of DGs and, at a late 
cosmic epoch, favor a nearly synchronous, dominant DG formation phase. 
Evidently, this question is complex.
Even rough quantitative statements would require reasonable assumptions on
galaxy statistics (e.g. mass, velocity and angular momentum distribution),
and the effects of SF and close-by interactions in a loose group of co-evolving DGs.
However, an exploration of this hypothesis appears to be of considerable
interest both in the context of XBCD evolution and in the light of the recent finding 
that a substantial fraction of late-type DGs in the Local Supercluster form 
galaxy binaries \cite{KarachentsevMakarov2008}. 

\section{Pairwise dwarf galaxy formation and evolution}
\label{sec:2}

\vspace*{-3mm}
The following discussion attempts to delineate with simple heuristic arguments
basic aspects of the star formation history (SFH) of DGs
undergoing recurrent mild interactions in a loose galaxy group. 
It is restricted to the sizeable population of late-type DGs in the field, 
well outside the higher-density cosmic network, and does not consider 
strong interactions or merging.
The discussion is based on merely two considerations. The first one is
related to the triggering of SF through long-distance (20--100 kpc) tidal
interactions \cite{Icke1985} 
(see e.g. \cite{Pustilnik2001} for an enlightening discussion on the Icke
mechanism as possible triggering agent of starbursts in BCDs).
The second one is based on the empirical fact that the 3D gas distribution 
of DGs is related to their mass ($M$) and luminosity 
(e.g. \cite{Lo1993,RoychowdhuryChengalur2010}): 
more massive DGs evolve within an oblate or even disklike gaseous 
reservoir ($\epsilon < \epsilon_{\rm c}$) whereas the gas halo of lower-mass DGs is
less oblate or nearly spherical ($\epsilon > \epsilon_{\rm c}$),
where $\epsilon$ denotes the gas vertical-to-radial extent and
$\epsilon_{\rm c}$ is a mean value for late-type DGs.
An implicit, perhaps not undoubted assumption is that the dependence of 
$\epsilon$ on $M$ settles early enough 
and holds throughout the DG evolution.
Note that the 3D gas geometry ($\sim\epsilon$) has been taken into 
account in several theoretical studies exploring the energy requirements for
galactic winds from isolated DGs hosting a central starburst
(e.g. \cite{SilichTenorioTagle2001} and references therein).
Some of the considerations below derive from these studies.

For the sake of illustration, one may consider the case of tidally triggered SF in
a binary system consisting of a higher- and lower-mass DG A and B with 
($M>M_{\rm c}$; $\epsilon < \epsilon_{\rm c}$) and ($M<M_{\rm c}$; $\epsilon >
\epsilon_{\rm c}$), respectively.
In system A, small-scale SF-driven outflows of hot gas can likely escape 
relatively unimpeded along the steep vertical density gradient of its gas
disk, causing a rather moderate damage to it and leaving a significant
fraction of the cold gas reservoir intact and susceptible to ensuing
self-propagating or tidally triggered SF. Most probably, recurrent weak 
interactions will therefore enhance the average SF efficiency of a 
higher-mass DG and accelerate the buildup of its stellar component.
The opposite will presumably be the case for the lower-mass, nearly spherical system B:
in the absence of a substantial vertical gas density gradient, and 
depending on the energy injection rate and overall geometry, bubbles of hot
gas may remain trapped within the ambient cold gas over their long cooling time 
or grow and merge leading to its large-scale disruption. 
In either case, even a minor episode of tidally induced 
SF may result in a long-lasting quenching of SF activities.
The same is obviously true for a pair of low-mass weakly interacting DGs (B+B).

It is straight forward to generalize the above considerations to an ensample of
DGs in a galaxy group spanning a range in $M$ and $\epsilon$. 
The essential cumulative effect of galaxy co-evolution in the specific context will
likely be the accelerated buildup of higher-mass DGs and the retarded 
evolution of lower-mass systems, resulting in a lower SF efficiency, higher 
present gas-mass fraction and younger \lwt\ for the latter, and {\it vice versa}.
Therefore, in the scenario sketchily delineated here, formation and co-evolution 
of DGs in loose pairs and groups provides a mechanism that contributes to galaxy
downsizing by amplifying inherent mass-related galaxy evolution biases. 
This hypothesis is consistent with a nearly synchronous evolution for
equal-mass galaxies and {\it vice versa}, with a decreasing \lwt\ with decreasing $M$,
and with the presence of a faint substrate of old stars in the lowest-mass
DGs, provided that they remain bound to the DG halo.

The above considerations additionally suggest that not only the local DG density, 
but also the larger-scale (a few 100 kpc $\equiv$ $l$) environmental density 
is a relevant parameter of DG evolution.
Since tidally induced SF episodes are probably tiny, brief and asynchronous, 
important pieces of evidence regarding the long-term impact of the environment on DG
evolution may evade detection in studies relying on the narrow visibility 
time-window of the H$\alpha$ emission line only.
Differential studies of the SFH for a representative
probe of field DGs over a volume $\sim l^3$ might thus be of considerable
interest. Further insights might emerge from searches 
for post-starburst DG candidates with e.g. LOFAR \cite{KleinPapaderos2008}.

Whether or not the gradual rise of SF in an ensemble of co-evolving DGs will
be succeeded by a dominant phase of galaxy formation at a later cosmic epoch,
as probably is the case for several XBCDs, is unclear. A conceivable hypothesis is
that only DGs with specific intrinsic and environmental properties can develop 
along such an extreme evolutionary pathway. 
Various lines of evidence suggest that the SFH of field DGs is largely
influenced by the central density $\rho_{\star}$ of their evolving stellar 
component (\cite{Papaderos2002-IZw18} and references therein; 
see also \cite{Loose1982} for early theoretical work exploring the effect 
of a dense stellar background on gas stability conditions).
This is also supported by the fact that the $\rho_{\star}$ of starbursting 
DGs (BCDs) is by typically an order of magnitude higher than that of
DGs undergoing quasi-continuous SF (dwarf irregulars) \cite{Papaderos1996b}.
However, since stars are being inherited the kinematics of the gas out of which they 
are born, $\rho_{\star}$ (hence, the compactness of a DG) depends, to some extent,
on the exchange of angular momentum between weakly interacting gaseous halos
in a co-evolving ensemble of DGs. Nature and nurture are tightly woven.

\vspace*{-3mm}
\section{Synopsis}
\label{sec:3}

\vspace*{-3mm}
This work is motivated by the fact that some of the extremely metal-poor 
young XBCD candidates known in the nearby universe reside in galaxy pairs with strikingly 
similar properties. This points to an intriguing synchronicity in
the formation history of such binary galaxies and leaves space for alternatives
to the standard hypothesis, according to which the rapid, starburst-driven 
evolution of XBCDs is triggered by recent chance encounters between two unrelated 
gas-rich dwarf galaxies (DGs).
An alternative hypothesis proposed here is that some XBCDs form and co-evolve as 
pairs or in groups of low-mass baryonic units, experiencing perpetual mild 
interactions throughout their evolution.
If so, a question that arises is whether this co-evolution process can,
under certain conditions, initially slow down the buildup of DGs and, at a late 
cosmic epoch, favor a nearly synchronous, dominant phase of galaxy formation, as
seen in XBCDs.
Based on simple heuristic arguments, we argue that the co-evolution scenario 
offers a tenable working hypothesis for the exploration of DG evolution in
the field. It provides a mechanism that likely amplifies galaxy
downsizing trends through the acceleration of the evolution of more massive
systems and the deceleration of the evolution of less massive ones.
This scenario is consistent with the presence of a faint substrate of ancient
stars in most XBCDs and with a nearly synchronous evolution of equal-mass DGs (and {\it vice versa}).
The discussion given here is admittedly over-simplified. Its main goal,
however, is not to provide conclusive answers but to highlight some of the
questions raised by XBCDs and stimulate theoretical work on these
enigmatic objects.

\begin{acknowledgement}
P. Papaderos is supported by Ciencia
2008 Contract, funded by FCT/MCTES (Portugal) and POPH/FSE (EC).
\end{acknowledgement}

\end{document}